\documentclass[a4paper,11pt]{article}
\usepackage{pos}
\usepackage{graphicx,amsmath,color}
\usepackage{relsize}
\usepackage{enumerate}
\usepackage{slashed}
\usepackage{multirow}
\usepackage[normalem]{ulem}

\def\g{\gamma}

\def\beq{\begin{equation}}
\def\eeq{\end{equation}}
\def\bea{\begin{eqnarray}}
\def\eea{\end{eqnarray}}

\def\bit{\begin{itemize}}
\def\eit{\end{itemize}}

\def\l{\left}
\def\r{\right}

\def\s{\sigma}

\def\baa{\begin{array}}
\def\eaa{\end{array}}

\def\simgt{\mathrel{\lower2.5pt\vbox{\lineskip=0pt\baselineskip=0pt
           \hbox{$>$}\hbox{$\sim$}}}}
\def\simlt{\mathrel{\lower2.5pt\vbox{\lineskip=0pt\baselineskip=0pt
           \hbox{$<$}\hbox{$\sim$}}}}
\newcommand{\vev}[1]{ \langle {#1} \rangle }

\def\bfc{\begin{figure}\begin{center}}
\def\efc{\end{center}\end{figure}}

\definecolor{chromeyellow}{rgb}{1.0, 0.65, 0.0}
\definecolor{darkcoral}{rgb}{0.8, 0.36, 0.27}
\definecolor{cadmiumgreen}{rgb}{0.0, 0.42, 0.24}

\title{Populating dark sectors with relativistic bubble walls}

\author*[a]{Miguel Vanvlasselaer}

\affiliation[a]{{Theoretische Natuurkunde and IIHE/ELEM, Vrije Universiteit Brussel,
\& The International Solvay Institutes, Pleinlaan 2, B-1050 Brussels, Belgium }}

\emailAdd{miguel.vanvlasselaer@vub.be}

\abstract{In this talk, we present a mechanism of Dark Matter  production during 
first order phase transitions and happening via the collision of the bubble wall and plasma quanta. 
We will first study the possibility that the dark matter is produced via a renormalisable operator. We will observe that in this context the DM can be much heavier than the scale of the phase transition and  has a large initial velocity, leading to the possibility of the DM
to be warm today. 
We will then turn to more realistic scenarios where the Dark Matter sector is secluded and its interaction with the visible sector (including the Standard Model) originates 
from dimension-five and dimension-six operators. In this regime, we also find that such DM is typically heavy and warm today.
We study separately the cases 
of weakly and strongly coupled dark sectors,
where, in the latter case, we focus on glueball DM, which turns out to have very distinct phenomenological properties. For completeness, we also systematically compute the Freeze-In production of the dark sector and compare it with the bubble-plasma DM abundances. All the analytical results are collected in a table presented in this paper. }

\FullConference{Proceedings of the Corfu Summer Institute 2024 "School and Workshops on Elementary Particle Physics and Gravity" (CORFU2024)\
12 - 26 May, and 25 August - 27 September, 2024\\
Corfu, Greece\\}

\begin{document}
\maketitle

\section{Introduction}

The mechanism leading to the abundance of Dark Matter (DM) in the early universe is still the subject of an intense research. One of the favorite candidate is the \emph{freeze-out} mechanism, which leads to the well-known WIMP miracle and drove a vast research program. The negative current results at direct detection experiments however motivates us to study different mechanisms for the production of the universe abundance of DM. 

In this respect, one intriguing candidate for such a mechanism are the First Order Phase Transitions (FOPTs),  where the transition occurs from a metastable to a more stable vacuum state, i.e. to a deeper minimum of the potential. Such transitions could have occurred in the early universe, as they are naturally present in a large variety of motivated BSM models like composite Higgs\cite{Pasechnik:2023hwv, Azatov:2020nbe,Frandsen:2023vhu, Reichert:2022naa,Fujikura:2023fbi}, extended Higgs sectors\cite{Delaunay:2007wb, Kurup:2017dzf, VonHarling:2017yew, Azatov:2019png, Ghosh:2020ipy,Aoki:2021oez,Badziak:2022ltm, Blasi:2022woz,Banerjee:2024qiu}, axion models\cite{DelleRose:2019pgi, VonHarling:2019gme}, dark Yang-Mills sectors\cite{Halverson:2020xpg,Morgante:2022zvc} and $B-L$ breaking sectors\cite{Jinno:2016knw, Addazi:2023ftv}.

A large part of their appeal comes from the role that they could play in relating apparently uncorrelated phenomena such as baryogenesis \cite{Kuzmin:1985mm, Shaposhnikov:1986jp,Nelson:1991ab,Carena:1996wj,Cline:2017jvp,Long:2017rdo,Bruggisser:2018mrt,Bruggisser:2018mus,Morrissey:2012db,Azatov:2021irb, Huang:2022vkf, ,Baldes:2021vyz, Chun:2023ezg}, production of primordial black holes\cite{10.1143/PTP.68.1979,Kawana:2021tde,Jung:2021mku,Gouttenoire:2023naa,Lewicki:2023ioy}, production of possibly observable gravitational waves (GWs)\cite{Witten:1984rs,Hogan_GW_1986,Kosowsky:1992vn,Kosowsky:1992rz,Kamionkowski:1993fg, Espinosa:2010hh} and finally,  the production of heavy dark matter\cite{Falkowski:2012fb, Baldes:2020kam,Hong:2020est, Azatov:2021ifm,Baldes:2021aph, Asadi:2021pwo, Lu:2022paj,Baldes:2022oev, Azatov:2022tii, Baldes:2023fsp,Kierkla:2022odc, Giudice:2024tcp,Gehrman:2023qjn}. This last possibility will be the main focus of this paper. We will show that bubble walls, when they reach ultra-relativistic velocity display a boundary which is sharp enough to pair produce heavy particles from the thermal particles in the plasma, as this was exemplified in \cite{Azatov:2020ufh}. We will mainly make two observations (see \cite{Azatov:2021ifm, Azatov:2024crd} for the original derivations): first those pairs of particles can be much more massive than the scale of the phase transition. This opens a new avenue to circumvent the well known  \emph{Griest-Kamionkowski (GK) bound} of $O(100)$ TeV\cite{PhysRevLett.64.615}. Secondly, the pair of heavy particles is produced in a boosted way, typically with velocities much higher than the thermal velocities, and so can constitute Warm Dark Matter (WDM), even if they are much heavier than the keV scale. 

\begin{figure}[!h]
\hspace{-0.9cm} 
\centering{
\includegraphics[scale=0.14]{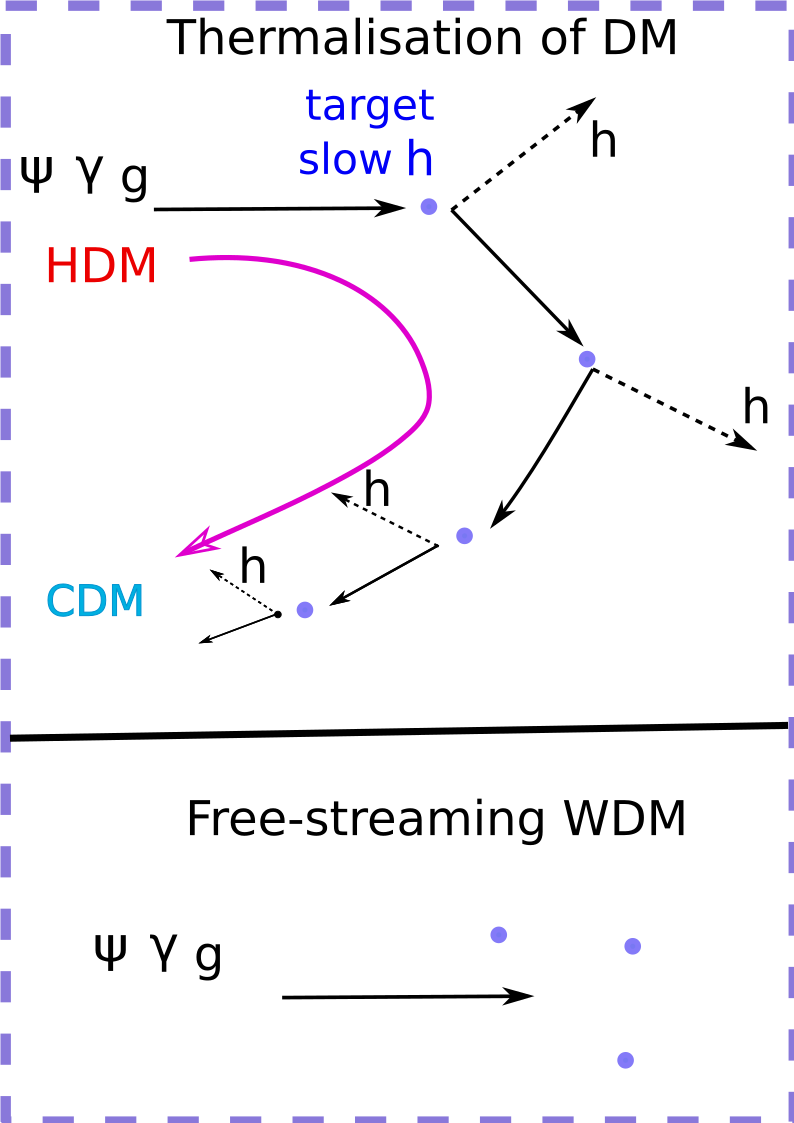}
\includegraphics[scale=0.23]{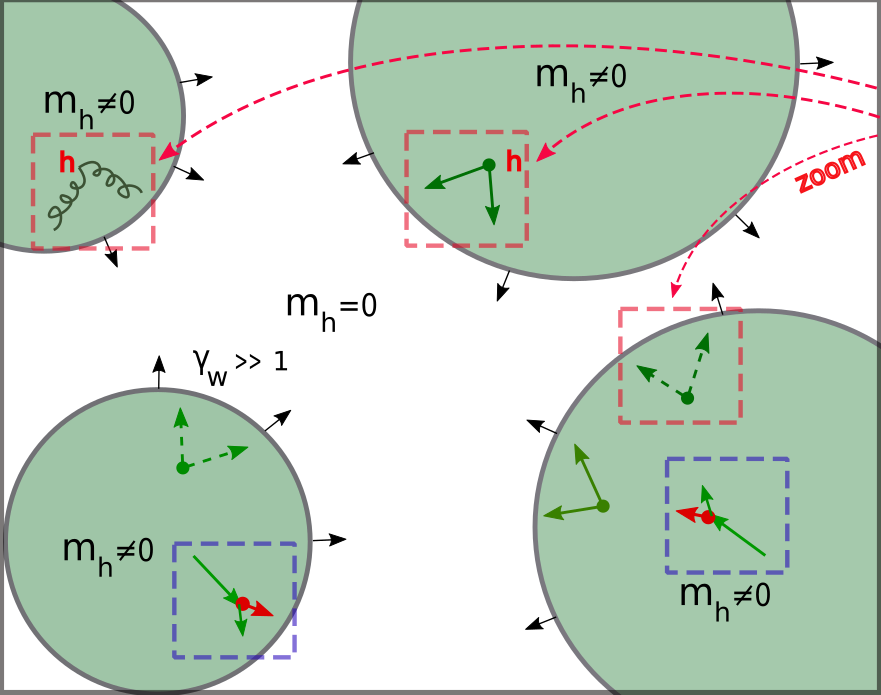}
\includegraphics[scale=0.23]{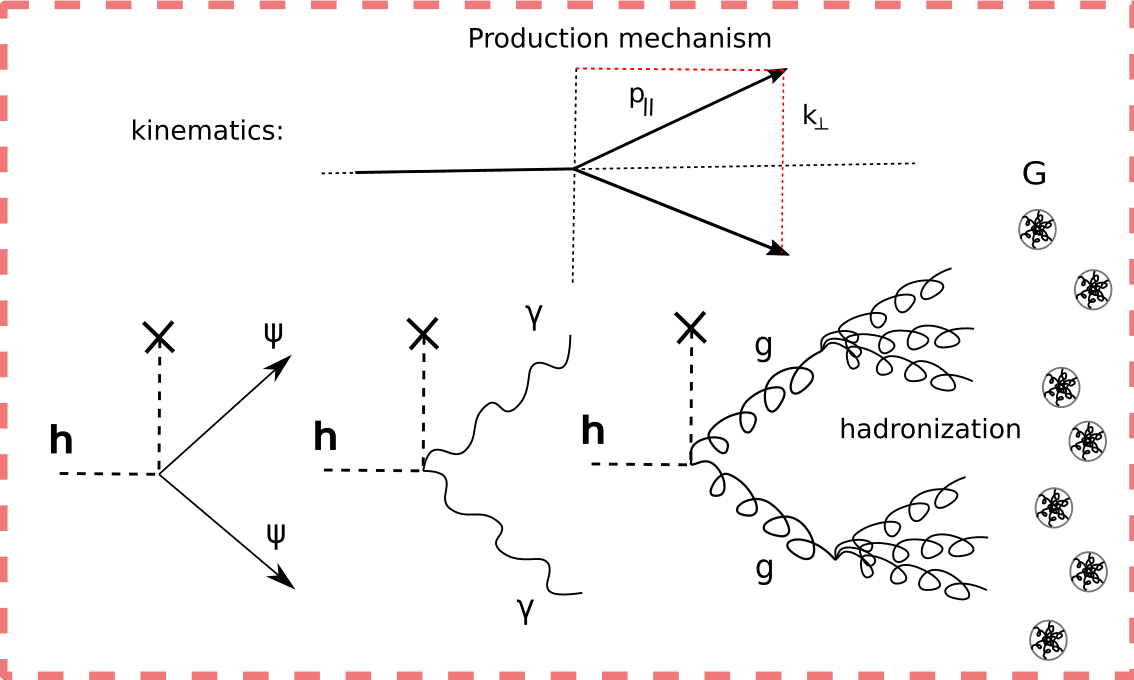}
}
 \caption{Schematic representation of the production of Dark Matter from bubbles. In the middle panel, we show how the expansion of bubble walls can produce very boosted scalars $\phi$, fermions $\psi$, vectors $\gamma$ or gluons $g$. In the right panel, the red rectangle, we present more in details the bubble-plasma production channels. The cross represents the interaction with the bubble wall which allows the DM production. In the left panel, in the blue rectangle, we show the interactions of the boosted produced particles with the thermalised plasma. The fast $\psi, \gamma, g$ can interact with the slow targets $h$ via $h(\psi, \gamma, g) \to h(\psi, \gamma, g)$ and cool down to usual CDM or free-stream and remain warm dark matter.  Adapted from \cite{Azatov:2024crd}.}
 \label{fig:sketch}
\end{figure}

We present a sketch of the mechanism in Fig.\ref{fig:sketch}, which displays the production via bubble wall expansion (Middle panel), the specific interaction between the plasma particle and the wall (in the right panel) and finally the subsequent interactions of the boosted particles with the thermalised plasma (in the left panel). 

\section{Dynamics of the phase transition and GW signal}

FOPTs proceed via the nucleation of bubbles of true phase inside of the false phase, at a temperature  defined by $H(T_{\rm nuc}) = \Gamma_{\rm nuc}$, where the rate $\Gamma_{\rm nuc}$ is the probability from unit of time and volume to nucleate a bubble. These bubbles then collide and percolate until the transition completes.  We conventionally define $\alpha_n$ as a parameter capturing the strength of the phase transition
\bea 
\alpha_n \equiv \frac{\Delta V}{\rho_{\rm rad}}  \, ,
\eea 
where $\Delta V$ is the energy release,  the difference of vacuum energy between the two phases.
\paragraph{Early stages}
Bubbles nucleate generically with a radius 
\bea 
R_{\rm nuc} \propto 1/T_{\rm nuc} \,. 
\eea 
The expansion of a bubble can proceed in two regimes: the bubble can reach a 
steady state motion at a given velocity, this is the \emph{terminal velocity regime}, or 
the bubble keeps accelerating until collision, in the \emph{runaway} regime. As long as the pressure from the release of energy 
is not balanced by the pressure from the plasma, the bubble keeps accelerating with a boost factor proportional to the radius of the bubble\cite{PhysRevD.45.3415,Ellis:2019oqb}
\bea
\label{eq:boost}
\gamma^{}_w (R)\approx  \frac{2 R}{3 R_{\rm nuc}}\l(1-\frac{\Delta \mathcal{P}}{\Delta V}\r) \approx  \frac{2 R}{3 R_{\rm nuc}} \,.
\eea 
In the runaway regime, the largest boost factor is naturally given by evaluating Eq.\eqref{eq:boost} the radius of the bubble at collision, $\gamma_w^{\rm ter} (R_\star)$: 
\bea
 R_\star\approx \frac{(8 \pi)^{1/3}v_w}{H[T_{\rm nuc}]\beta(T_{\rm nuc})},~~~\beta(T)= T\frac{d}{dT}\l(\frac{S_3}{T}\r) \, , \qquad \gamma^{\rm coll
}_w \sim \frac{2 \sqrt{10} M_{\rm pl} T_{\rm nuc}}{ \pi^{2/3}\sqrt{8\pi} \sqrt{g_\star} \beta T_{\rm reh}^2} \approx 0.06\frac{ M_{\rm pl} T_{\rm nuc}}{ \beta T_{\rm reh}^2} \, ,
\eea
where $R_\star$ is an estimate for the bubble size at collision and $\beta$ the inverse dimensionless duration parameter of the transition.   

\paragraph{Production of particles and pressure}
The pressure from particles coupling to $h$ might terminate the acceleration long before $\gamma_w^{\rm collision}$ is reached\cite{Dine:1992wr,Liu:1992tn,Moore:1995ua,Moore:1995si,Dorsch:2018pat,Laurent:2022jrs,Jiang:2022btc,Konstandin:2010dm, BarrosoMancha:2020fay,Balaji:2020yrx, Wang:2022txy,Krajewski:2023clt, Sanchez-Garitaonandia:2023zqz,Bodeker:2009qy, Bodeker:2017cim, Azatov:2020ufh,Gouttenoire:2021kjv,Ai:2023suz,Azatov:2023xem,Ai:2021kak,Ai:2023see, Ai:2024shx, Azatov:2024auq,Barni:2024lkj}. In the ballistic approximation, the computation of the terminal velocities \cite{Bodeker:2009qy, Bodeker:2017cim, Azatov:2020ufh,Gouttenoire:2021kjv,Azatov:2023xem} amounts to comparing the release of energy $\Delta V$ with the plasma pressure in the relativistic regime $ \mathcal{P} (\gamma_w)$, 
$
\Delta V \approx  \sum \mathcal{P}(\gamma^{\rm ter}_w) 
$,
where $\sum \mathcal{P}(\gamma^{\rm ter}_w)$ is the sum of the different source of pressure. In this approximation, one neglects the scatterings among particles inside the bubble wall, which are expected to reduce the pressure\cite{Ai:2024uyw} and the pressure is due to the interactions inducing an exchange of momentum from the plasma to the bubble wall. It reads
\bea 
\label{eq:intuitive_picture}
\mathcal{P}^{\gamma_w \to \infty} \approx \sum_{ij}\underbrace{\frac{p_z}{p_0} n_i}_{\text{flux}} \times \underbrace{P_{i \to j}}_{\text{probability $i \to j $}} \times \underbrace{\Delta p_{i \to j}}_{\text{exchange of momentum $i \to j$}}
\eea 
where the first factor is the incoming flux of particle species $i$ entering into the wall and having a transition $i \to j$ , i.e. to state $j$, with an associated loss of momentum $\Delta p_{i \to j} \equiv p_i - p_j$. This loss of momentum of the plasma is transmitted to the wall, which is felt by the wall as a pressure. 

\begin{figure}
    \centering
    \includegraphics[width=\textwidth]{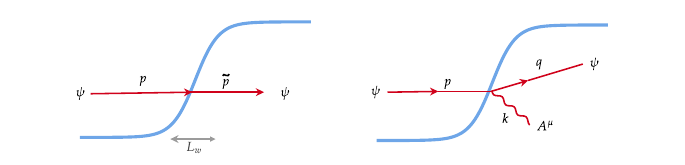}
    \caption{Schematic representation of the processes responsible for the friction at LO (Left) and NLO (Right).  The emission of vectors with changing mass is generally the dominant process.  Adapted from \cite{Azatov:2023xem}. }
    \label{fig:cartoon}
\end{figure}

 The first contribution to the pressure is the pressure from the particles coupling to the BSM Higgses $h$ and consequently gaining mass durign the transition\cite{Bodeker:2009qy}, which are presented on the Left panel of Fig.\eqref{fig:cartoon}, 
\begin{align}
\label{eq:mass_gain}
\mathcal{P}_{h} \approx g_h\frac{m_h^2T^2_{\rm nuc}}{24} \qquad \qquad 
\mathcal{P}_{i} \approx c_ig_i\frac{m_i^2T^2_{\rm nuc}}{24} \, ,
\end{align} 
where $m_h$ is the mass of the BSM Higgs in the broken phase, $c_i = 1(1/2)$ for bosons(fermions). On the other hand, if the BSM Higgs is gauged, the emission of soft transverse gauge bosons\cite{Bodeker:2017cim, Gouttenoire:2021kjv} and longitudinal gauge bosons\cite{Azatov:2023xem}, presented on Right panel of Fig.\eqref{fig:cartoon}, would induce a pressure
\bea 
\label{frictiongauge}
\mathcal{P}_{g} \propto \frac{g^3}{16 \pi^2} \gamma_w T_{\rm nuc}^3 v \qquad \text{(emission of soft bosons: model-dependent)} \, ,
\eea 
preventing runaway. In what follows we will assume that phase transition sector is not gauged such that $\mathcal{P}_{g} \to 0$.

\paragraph{Gravitational wave signal from FOPTs}

\begin{figure}[!h]
\hspace{-0.9cm} 
\centering{
\includegraphics[scale=0.2]{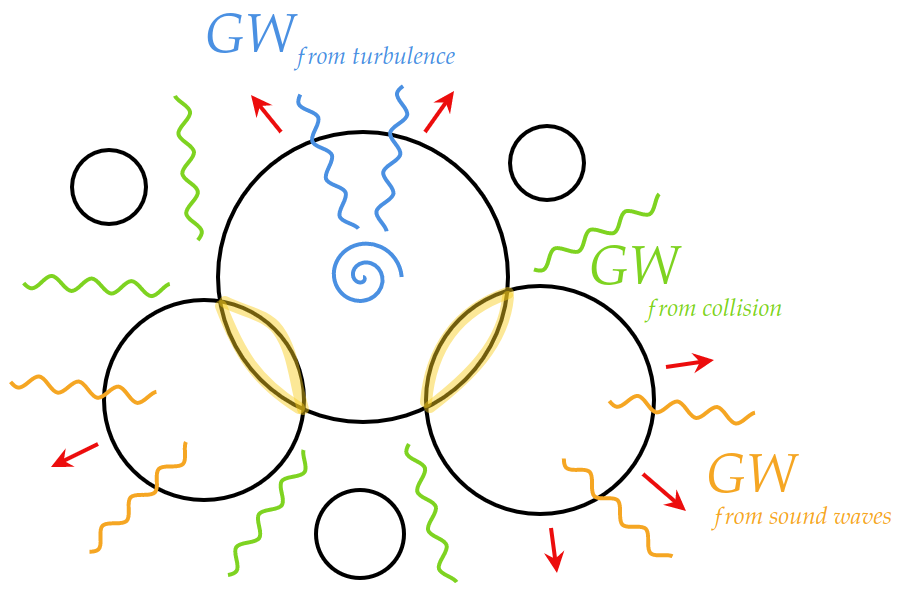}
\includegraphics[scale=0.5]{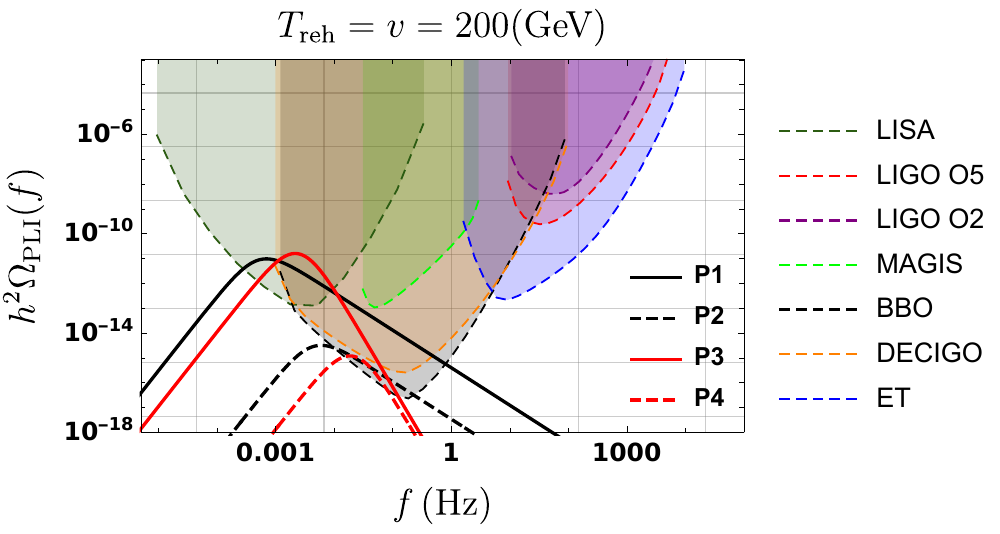}
}
 \caption{Left: Schematic presentation of the different contributions to the GW spectrum from FOPTs. Credit: \emph{Giulio Barni}. Right: GW signal against the predicted sensitivities of GW observers. {\bf Left}-GW signal with $v = T_{\text{reh}} = 200$ GeV for four benchmark points in four different regimes: P1 (runaway $ \alpha = 1, \beta = 100$), P2 (runaway $ \alpha = 0.1, \beta = 1000$), P3 (terminal velocity $ \alpha = 1, \beta = 100$), P4 (terminal velocity $ \alpha = 0.1, \beta = 1000$). Sensitivities are obtained from \cite{Moore:2014lga,Aasi:2013wya,TheLIGOScientific:2014jea,Cornish:2018dyw,Graham:2017pmn,Yagi:2011yu,Yagi:2013du,Sathyaprakash:2012jk}. Adapted from \cite{Azatov:2021ifm}.  }
 \label{fig:GW_from_FOPTs}
\end{figure}

FOPTs source Gravitational Waves (GWs) via three different mechanisms (see \cite{Caprini:2015zlo} and references therein): bubble wall collision (green contribution on Fig.\ref{fig:GW_from_FOPTs}) which occurs on a timescale set by $1/R_{\star}$, sound wave propagation, which source GWs until the onset of non-linearities in their hydrodynamics (orange contribution on Fig.\ref{fig:GW_from_FOPTs}) and finally the turbulence (in blue on Fig.\ref{fig:GW_from_FOPTs}). The magnitude of the gravitational wave signal would typically scale like 
 $\Omega_{\rm GW} \propto v_w (\alpha/(1+\alpha))^2/\beta^2$, which is maximised for large energy budget $\alpha_n$ and small $\beta$, long phase transition duration. Such conditions can typically by fulfilled if the transition sector is controlled by a broken \emph{conformal symmetry}\cite{Konstandin:2011dr, Baratella:2018pxi, Azatov:2019png}. Since $\mathcal{P}_h \propto \alpha_n^{-1/2}$, large $\alpha_n$ are also the optimal condition for fast bubbles and for the copious production of heavy dark matter, as we will observe now.

\section{Production of dark matter via renormalisable operator}

We now turn to the production of heavy states from the bubble wall. The simplest interaction that we can consider is the \emph{renormalisable} interaction $\lambda \phi^2 h^2$, as it was initially proposed in \cite{Azatov:2021ifm} and then further studied in \cite{Baldes:2022oev}. We thus consider the Lagrangian 
 \bea 
 \mathcal{L} = \frac{1}{2}(\partial_\mu \phi)^2 - \frac{1}{2} M_\phi^2 \phi^2 - \frac{\lambda}{2}h^2 \phi^2  \, .
 \eea

\paragraph{Abundance produced}
During the phase transition, the BSM Higgs field $h \to h+v$, induces a three-leg vertex in the Lagrangian  of the form
\bea 
\mathcal{L} \subset \lambda v h \phi \phi \quad,
\eea
allowing for splittings $h \to \phi \phi$. It is clear that this transition would be forbidden in vacuum due to conservation of momentum and only occurs thanks to the bubble wall presence, which breaks Lorentz invariance, leading to the non-conservation of $z$-momentum. Using the WKB approximation, the transition from light to heavy states $h \to \phi\phi$ has a probability\cite{Azatov:2021ifm}
\bea
\label{eq:prod_scalar}
P_{h \to \phi^2} \simeq \bigg(\frac{\lambda v}{M_\phi}\bigg)^2\frac{1}{48\pi^2} \Theta \l( p_0-\frac{2M_\phi^2}{v} \r).
\eea
In Eq.\eqref{eq:prod_scalar}, $L_w$ is the width of the wall, which is approximately  $L_w
 \sim 1/v$, with $v \ll M_\phi$, and $\Delta p_z \equiv p^h_z - p^\phi_{z,b}-p^\phi_{z,c} \approx 2M_\phi^2/p^h_z$ is the difference of momenta 
 between final- and initial-state particles in the direction 
 orthogonal to the wall.  The $\Theta( p_0-2M_\phi^2/v)$-function comes from the requirement that the transition is
  \emph{non-adiabatic}, that is to say, that there is enough energy in the center of mass collision between a quanta $h$ from the wall and the plasma quanta $h$: $s \approx 2 p_0v \gg 4M_\phi^2$  (see Appendix A of \cite{Azatov:2021ifm} and Appendix H of \cite{Azatov:2024auq} for more details about this computation). 
  The individual emitted particles are very boosted in the plasma frame\cite{Azatov:2021ifm}, with average energy in the plasma frame
$
\bar{E}_{\phi, \text{plasma}} \approx   M_\phi^2/2T_{\rm nuc}$. Behind the bubble, a non-thermal  abundance of $\phi$ accumulates,
\bea
n_\phi^{\text{BE, PF}} \approx 
 \frac{2\lambda^2 v^2}{48\pi^2 M_{\phi}^2 \gamma_w v_w}  \int \frac{d^3p}{(2\pi)^3}  \times f_h (p,T_\text{nuc})\Theta ( p_z- 2M_\phi^2/v),
\label{eq:density_1}
\eea 
where the subscript PF means that the quantity has been evaluated in the plasma frame. 
 $v_w = \sqrt{1-1/\gamma_w^2}$ is the velocity of the wall, and $f_h(p)$ is the equilibrium thermal distribution of $h$ outside of the bubble.  
Since $h$ is in thermal equilibrium with the bath at temperature $T_{\rm nuc}$ it can be described by a Boltzmann distribution  $f_h (p) \approx e^{-{\gamma_w}(E_h - v_wp^h_z)/T_\text{nuc}}$. We can thus perform the integral in Eq. \eqref{eq:density_1}
\bea
n_\phi^{\text{BE}} &=&  \frac{T_\text{nuc}^3}{24\pi^2} \frac{\lambda^2 v^2}{\pi^2 M_\phi^2}  e^{-  \frac{M_\phi^2}{vT_\text{nuc} \gamma_w }}  + \mathcal{O}(1/\gamma_w) \quad ,
\label{eq:density_f}
\eea
where we used that $\gamma_w (1-v_w) = \gamma_w - \sqrt{\gamma_w^2 - 1} \to \frac{1}{2\gamma_w}$ and considered  the limit of fast walls. The factor $e^{-  M_\phi^2/(vT_\text{nuc} \gamma_w )}$ is a consequence of $\Theta ( p_0- 2M_\phi^2/v)$ in the the Eq. \eqref{eq:density_1} and this implies that in the non-adiabatic limit, $
 \gamma_w > M_\phi^2/vT_{\text{nuc}} , $
 the exponential goes to one and the density becomes independent of the velocity of the wall $v_w$, while in the adiabatic limit $\gamma_w < M_\phi^2/vT_{\text{nuc}}$ the production becomes exponentially suppressed, as expected.

After the bubbles collided, the transition completed and plasma thermalised to the \emph{reheating temperature} $T_{\rm reh}$, that can be estimated via conservation of energy to be
$
T_{\rm reh} \approx (1+\alpha_{\rm nuc})^{1/4} T_{\rm nuc} \approx v  \,.  
$
 This reheating captures the fact that the phase transition injects entropy in the plasma. After this entropy injection, the final relic abundance today becomes 
\bea
\Omega^{\text{today}}_{\phi,\text{BE}}h^2 \approx {2.7}\times 10^5  \times \bigg(\frac{1}{g_{\star S}(T_{\text{reh}})}\bigg) \bigg(\frac{\lambda^2 v}{ M_\phi}\bigg)\bigg(\frac{v}{\text{GeV}}\bigg)\bigg(\frac{T_\text{nuc}}{T_{\text{reh}}}\bigg)^3e^{-  \frac{M_\phi^2}{vT_\text{nuc} \gamma_w }}.
\label{eq:relic_ab}
\eea
where $g_{\star S}(T)$ is the relativistic degrees of freedom for entropy. 

At this point, two avenues are possible: the first is to assume that $M_\phi \gg 20 T_R$, where $T_R$ is the highest temperature after inflation and the heavy DM is never produced thermally, nor via freeze-in. In such case, one can fit the observed abundance of DM directly from Eq.\eqref{eq:relic_ab} and see that, for example, the point with $\lambda =1$, mild supercooling, $M_\phi \sim 10^8$ GeV and $v \sim 10^2$ GeV would typically produce the required amount of DM. The second avenue is to assume that the candidate DM $\phi$ thermalised in the early universe and compare the abundance from FO (freeze-out) and BE (bubble expansion). After thermal inflation, the FO abundance receives a further $\big(\frac{T_\text{nuc}}
{T_{\text{reh}}}\big)^3$ suppression factor with respect to usual cosmology evolution. Summing both  FO and BE contributions the total relic abundance will be approximately given by (we are assuming $M_\phi \gtrsim 20 T_{\rm reh}$)
\bea
\Omega^{\text{today}}_{\phi, \text{tot}}h^2 \approx \l(\frac{T_\text{nuc}}
{T_{\text{reh}}}\r)^3 \times \bigg[\underbrace{0.1\times \bigg(\frac{0.03}{\lambda}\bigg)^2 \bigg(\frac{M_\phi}{100 \text{ GeV}}\bigg)^2}_\text{FO} + \underbrace{5\times 10^3 \times \lambda^2\frac{ v}{M_\phi}\bigg(\frac{v}{\text{GeV}}\bigg)}_\text{BE} \bigg].
\label{eq:total_ab_dil}
\eea

On the other hand, if $T_{\rm reh}\gtrsim 1/20 M_\phi$, the contribution from FO after the PT can become important.  Assuming an instantaneous reheating after bubble collision and negligible non-thermal production of $\phi$ via bubble collision \cite{Falkowski:2012fb}, the additional contribution from thermal production is given by\cite{Azatov:2021ifm} 
\bea
\delta  \Omega_{\phi, \rm tot}^{\rm today} \sim M_\phi\left.\frac{\vev{\s_{\phi\phi} v_{\rm rel}}n_{\rm eq}^2}{H g_{\star S}(T)T^3}  \right|_{ T=C' T_{\rm reh}}\times  \frac{g_{\star S0}T_0^3}{\rho_c} .
\label{eq:thermalprod}
\eea

\begin{figure}
\centering
\includegraphics[scale=0.27]{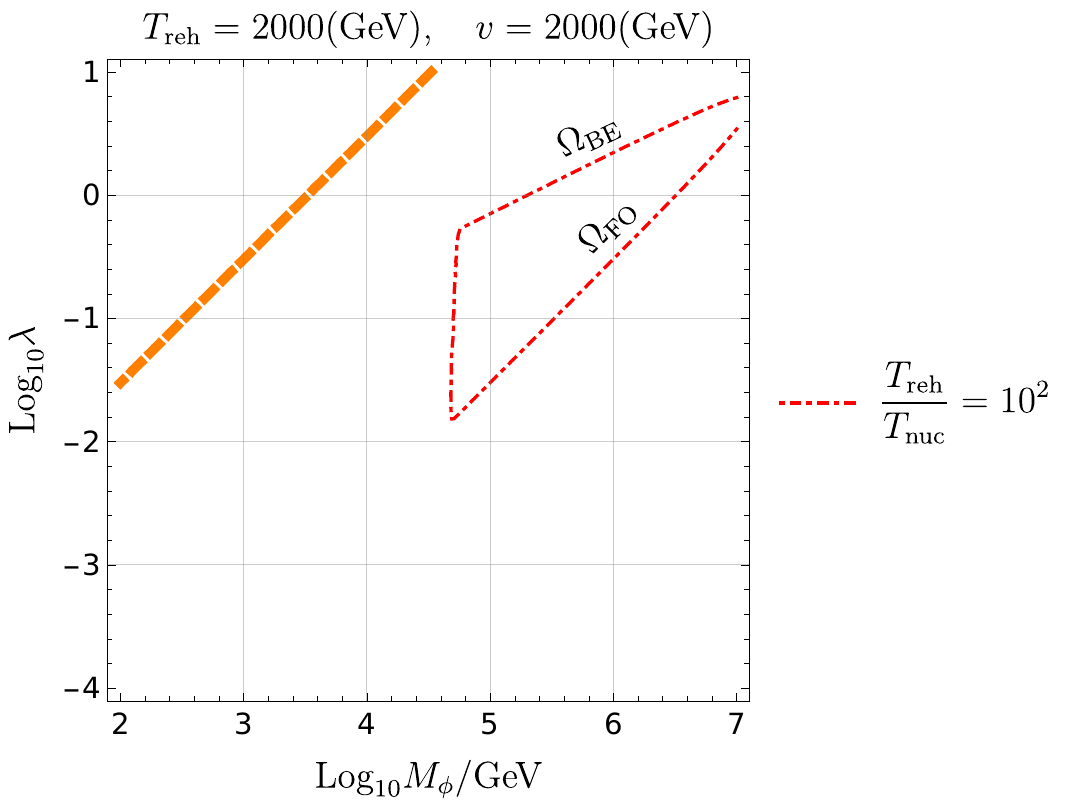}
\includegraphics[scale=0.37]{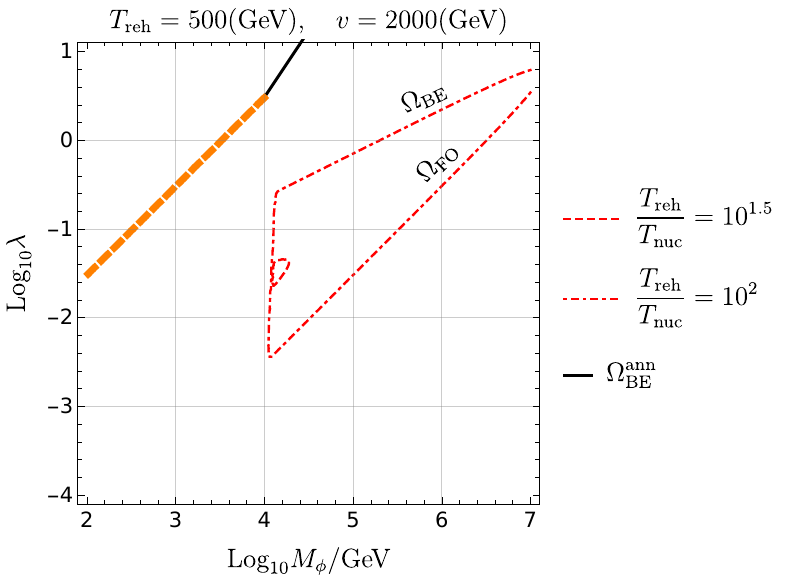}
\includegraphics[scale=0.37]{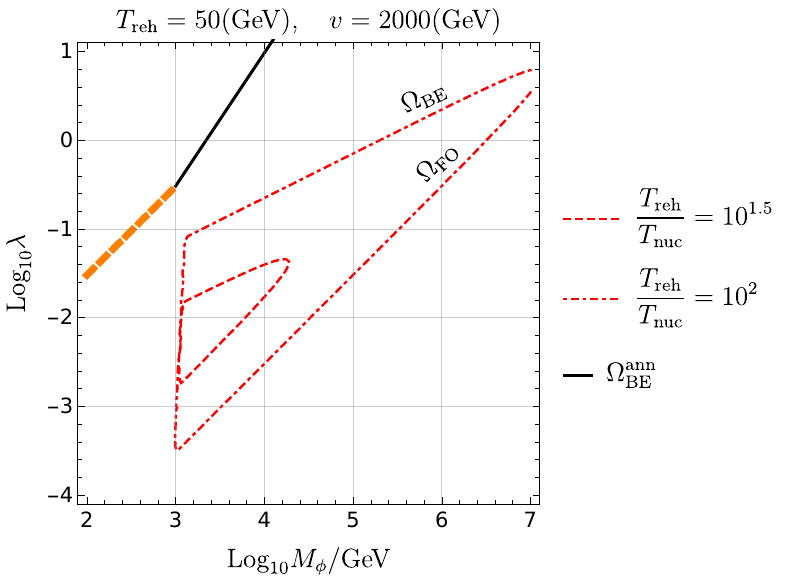}
\caption{Values of $M_\phi$ and $\lambda$ providing the observed DM relic abundance today if $h$ is a dark higgs, for various values of supercooling $\frac{T_{\text{reh}}}{T_{\text{nuc}}} = (10, 10^{1.5},10^2)$, $v =  2000$ GeV. The Red lines show the contributions from FO and BE providing the observed DM abundance and that do not undergo annihilation after the transition. The black line is the result of DM annihilation via $\phi \phi \to h h$. When $M_\phi < 20 T_{\text{reh}}$, the DM candidate comes back to equilibrium after the transition the usual FO abundance applies. We also assumed runaway walls. Figure adapted from \cite{Azatov:2021ifm}.  }
\label{Fig:regions3}
\end{figure}

Putting the contributions from Eq.\eqref{eq:thermalprod} and Eq.\eqref{eq:total_ab_dil} together, one can study the region of the parameter space where the observed abundance of DM is fulfilled. We present our result on Fig.\ref{Fig:regions3}, for various valyus of $v$ and $T_{\rm reh}$, involving often a long supercooling.

\paragraph{Spectrum and velocity of the emitted DM} Due to the large initial boost of the heavy particles, the produced particles can have a significant  \emph{free-streaming} (FS) length $L_{\rm FS}$\cite{Baldes:2022oev}
\bea
L_{\rm FS}=
\int^{\infty}_{z_{\rm eq}} dz \frac{1}{H}\frac{{V_{\rm eq} \frac{1+z}{1+z_{\rm eq}}}}{\sqrt{\Big(V_{\rm eq}\frac{1+z}{1+z_{\rm eq}}\Big)^2+1}},
\eea 
where we defined $ V_{\rm eq}\equiv V(t_{\rm eq})$ as the average velocity of the DM at matter-radiation equality and $z$ is the redshift. The strongest constraint for the DM free-streaming comes from Lyman-$\alpha$ for the DM free-streaming length $L_{\rm FS}\lesssim 0.059{\rm Mpc}$, which is recast from sterile neutrino mass bound, $5.3$ keV\cite{Bode:2000gq,Viel:2005qj,Irsic:2017ixq}. 
This leads to the following bound on the average velocity at matter-radiation equality, 
\bea 
V_{\rm eq} 
\lesssim 4.2\times 10^{-5}
\qquad \text{(Lyman-$\alpha$ bound)} \, , 
\label{eq:LyAlphaBoundNow}
\eea 

The DM particles which have such non-negligible velocities are denoted in the literature as warm (WDM), with  typical velocities
$
V^{\rm WDM}_{\rm eq}\sim 10^{-5}.
$
In the thermal mechanisms\cite{Colombi:1995ze} for the production of WDM, 
the DM candidate mass has to be small, around keV mass scale. For the case of bubble wall production, assuming that the DM  is free-streaming after the phase transition\footnote{This assumption is verified in \cite{Baldes:2022oev, Azatov:2024crd}}, we can easily observe that 
 \bea 
\label{eq:velocity_eq}
V_{\rm eq} \approx \bigg(\frac{g_{\star,s}(T_{\rm eq})}{g_{\star,s}(T_{\rm reh})}\bigg)^{1/3}\frac{T_{\rm eq} p^i_{\rm DM}}{T_{\rm reh} M_{\rm DM}} \approx 0.3 \frac{T_{\rm eq} \bar{E}_\phi}{T_{\rm reh} M_{\phi}}  \approx   10^{-10} \frac{ \text{GeV} \times M_{\phi}}{T_{\rm reh} T_{\rm nuc}} , 
\eea
where $T_{\rm eq}$. Thus the DM produced during the bubble expansion will be warm if 
$
10^5 v T_{\rm nuc} \sim  M_\phi \times \text{  GeV} \, .
$
Very interestingly, this indicates a viable parameter space for explaining the observed abundance of DM being heavy and warm in a range roughly $v \sim \mathcal{O}(100) \text{GeV}, M_\phi \sim 10^{(8-9)} $ GeV and mild supercooling\cite{Baldes:2022oev}.

\section{Non-renormalisable production of dark matter}

\begin{table}
    \centering
    \renewcommand{\arraystretch}{2}
    \begin{tabular}{c|c|c|c|}
    & $h^2\psi \bar{\psi}/\Lambda$ & $h^2FF/\Lambda^2$ & $ h^2GG/\Lambda^2$ \\
    \hline
         $P_{1\to 2}$ &  $\frac{v^2}{\Lambda^2} \left(\log\frac{2p_0v}{M^2}-4/3\right)$ & \multicolumn{2}{c|}{ $\frac{v^3p_0}{\Lambda^4}$}  \\ 
         \hline
$\Omega^{\text{}}_{\text{BE}}h^2$ & \multicolumn{2}{c|}{ $ \frac{M}{\rm GeV}P_{1\to 2} \alpha_n^{3/4} e^{-\frac{M^2}{vT_{\rm nuc }\g_w} } $  }& $ \frac{M}{\rm GeV}\left(\g_w P_{1\to 2}\right)^{3/4} \alpha_n^{3/4} $
 \\
         \hline
$\Omega^{\text{}}_{\text{FI}}h^2$ & 
$
 60 \frac{M^2 M_{\rm pl}}{\Lambda^2 \times \text{GeV}}  \left(\frac{ T_{\rm reh}}{M}\right)^{3/2} e^{-\frac{2M}{T_{\rm reh}}}
$
&
$
10^2\frac{M^3 M_{\rm pl} }{ \Lambda^4} \sqrt{\frac{M}{T_{\rm reh}}} e^{-\frac{2M}{T_{\rm reh}}}
$
&  
\\ \hline
$\bar{E}$ &  $ 
\frac{L_w^{-1} \gamma_w}{3\log ( \gamma_w T/M^2 L_w) -5.92}$  & \multicolumn{2}{c|}{ $ 0.16\g_w L_w^{-1}$ }
\\ \hline
$V_{\rm eq}$ & 
\multicolumn{2}{c|}
 { $\bar{E} \frac{{\rm GeV }}{T_{\rm reh} M}$ ~~~if $\Lambda^2 > M_{\rm pl}T_{\rm reh}$    } & No Free Streaming
\\\hline
\color{black}
$\mathcal{P}_{1 \to 2}$ & 
$  \frac{1}{8\pi^4}\frac{\gamma_w v^3 T^3 }{\Lambda^2}$  & \multicolumn{2}{c|}{ $
 \frac{1}{2\pi^4}  \frac{\gamma_w^2 v^4 T^4}{\Lambda^4}$ } \\
\hline
$M/\text{GeV}$ & $[10^6-10^{14}]$ & $[10^6-10^{14}]$ &  $[1-10^6]$\\
\hline
$v/\text{GeV}$ & $[10^3-10^{12}]$ & $[10^2-10^{14}]$ &  $[10^2-10^{14.5}]$\\
\hline
    \end{tabular}
    \caption{In this table, we present a summary of the important results. In these expressions, $\gamma_w$ is the boost factor for the wall to the plasma, $v$ is the VEV of the broken symmetry, $P_{1 \to 2}$ is the probability of the splitting, $\mathcal{P}$ is the averaged pressure on the bubble wall from the reaction, $V_{\rm eq}$ is the velocity of the DM at radiation-matter equality, $\bar E$ is the average energy of the DM particle \emph{at production}. The subscripts BE and FI mean \emph{bubble expansion} and \emph{freeze-in} abundance, respectively. In the last two rows, we report the attainable DM mass and symmetry breaking scale, \emph{assuming} that the DM  abundance via bubble expansion matches the observation. }
    \label{tab:Summary_tab}
\end{table}

More realistically, the dark sector are only coupled to the SM via higher dimension operators, in which case they are \emph{secluded}. We now turn to the evaluation of the production from these \emph{non-renormalisable} operators. We study first, fermionic DM $\psi$, which carries conserved quantum number responsible for its stability, coupled to the thermalised sector $h$ via the following non-renormalisable operator 
\bea 
\label{eq:fermion_op}
 \mathcal{L} \supset \frac{h^2 \psi \bar \psi}{\Lambda} \; . 
\eea 

\begin{figure}[!h]
\hspace{-0.9cm} 

\includegraphics[scale=0.6]{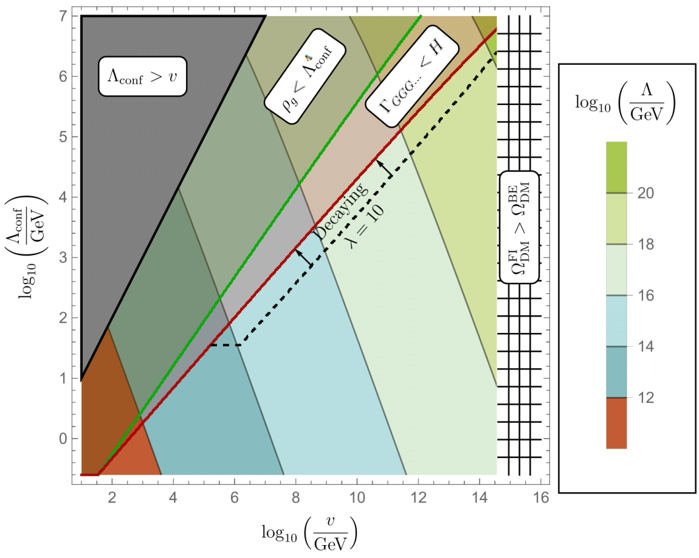}
 \caption{Parameter space where the production of Glueballs can fit the observed amont of dark matter. We show in black dashed the region where the Glueballs are decaying. Adapted from \cite{Azatov:2021ifm}.}
 \label{fig:para_spaceGG}
\end{figure}

Then we turn to a dark photon coupled to the scalar field via
\bea 
\frac{h^2 F_{\mu \nu} F^{\mu \nu}}{\Lambda^2} \, ,
\eea 
where $F_{\mu \nu}$ is the field strength of a dark vector $\gamma$. This vector DM can be made stable by assuming a sufficiently small mixing with the SM photon. Finally, we investigate the production of dark gluons which leads to Glueballs dark matter via 
\bea 
\frac{h^2 G_{\mu \nu}G^{\mu \nu}}{\Lambda^2} \,. 
\eea 
 While the analysis of the production of fermions and dark photons proceed essentially in similar way than the production of scalar, and we do not reproduce the analysis here (see \cite{Azatov:2024crd}), the Glueballs production include several subtleties. The gluons produced can directly hadronise, or produce a thermal gas of gluon.  We find two conditions for reaching a state of gluon plasma after the phase transition: First, one should transfer sufficient amount of energy to the dark sector, which amount to requiring
\bea 
\label{eq:qg_enough_E}
\rho_g  \approx  C_g (N^2 -1)\frac{2\g^2_w v^4T_{\rm nuc}^4}{(\pi\Lambda)^4} >\Lambda_{\rm conf}^4 \, .
\eea 
This condition is satisfied outside of the green region of Fig.\ref{fig:para_spaceGG}. On the other hand,  if the initial density of glueballs is very small, a gluon plasma will not be formed. To estimate the minimal necessary density, we require that the scattering rate of glueballs is larger than the Hubble expansion rate:
$
\Gamma_{G G \to GG}  \sim \Gamma_{GGG ...}   > H\, ,
$
where the rate $\Gamma_{GGG ...}$ is the total  rate of the processes in which three or more glueballs are produced in $GG$ collisions.  This condition is satisfied outside of the red region of Fig.\ref{fig:para_spaceGG}.

In the rest of the analysis, we will stick to the regime in which a gluon plasma is produced after the bubble wall expansion. In this regime, the calculation of the glueball relic abundance proceeds in a standard way \cite{Carlson:1992fn,Hochberg:2014dra,Forestell:2016qhc}.  
The relic abundance of Glueball DM today is given by
\bea
&&\frac{(\Omega h^2)_G}{(\Omega h^2)_{\rm DM}}\approx 0.056 (N^2-1)\l(\frac{B}{10^{-12}}\r)^{3/4}
\l(\frac{\Lambda_{\rm conf}}{\rm GeV}\r)W\l[2.1\frac{N^2-1}{N^{18/5}} B^{3/10}\l(\frac{M_{\rm pl}}{\Lambda_{\rm conf}}\r)^{3/5}\r]^{-1} \; ,
\\
&& B\equiv \frac{T_{g}^4}{T_{\rm SM}^4}= \frac{g_{\star} \rho_{g}}{2(N^2-1)\rho_{\rm SM}} = \frac{ 30\rho_{g}}{2\pi^2(N^2-1)T_{\rm SM}^4} =\frac{30C_g}{\pi^6}\frac{ \gamma_w^2  v^4}{ \Lambda^4}\bigg(\frac{T_{\rm nuc}}{T_{\rm reh}}\bigg)^{4} \, ,
\eea
where the function $W$ is the inverse function of $x e^x$.

 As the glueballs are not naturally protected by a symmetry, they will have a natural decay channel to a pair of $h$ via dimension six operator. To avoid this fast decay, we require that $m_h
 \sim v> M_G \sim 5\Lambda_{\rm conf}$ \cite{Curtin:2022tou}, so that the glueballs cannot directly decay to $h$.  Other decay products involve the SM particles and arise from dimension six operators and are more suppressed. We show in dashed lines on Fig.\ref{fig:para_spaceGG} the region of the parameter space which would display decaying DM.

\paragraph{EFT breakdown}
When the energy in the center of mass $s_{\rm prod} \approx 2\gamma_w vT_{\rm nuc} > \Lambda^2$, the EFT breaks down and we expect the production mechanism to become dependent on the explicit UV completion of the model. For this reason, we will require now that 
 \bea
 s_{\rm prod} \approx 2p_0 v \approx 2\gamma_w vT_{\rm nuc}< \Lambda^2 \qquad \text{(EFT validity condition)} \, . 
 \eea 

\begin{figure}[!h]
\hspace{-0.9cm} 
\centering{
\includegraphics[scale=0.5]{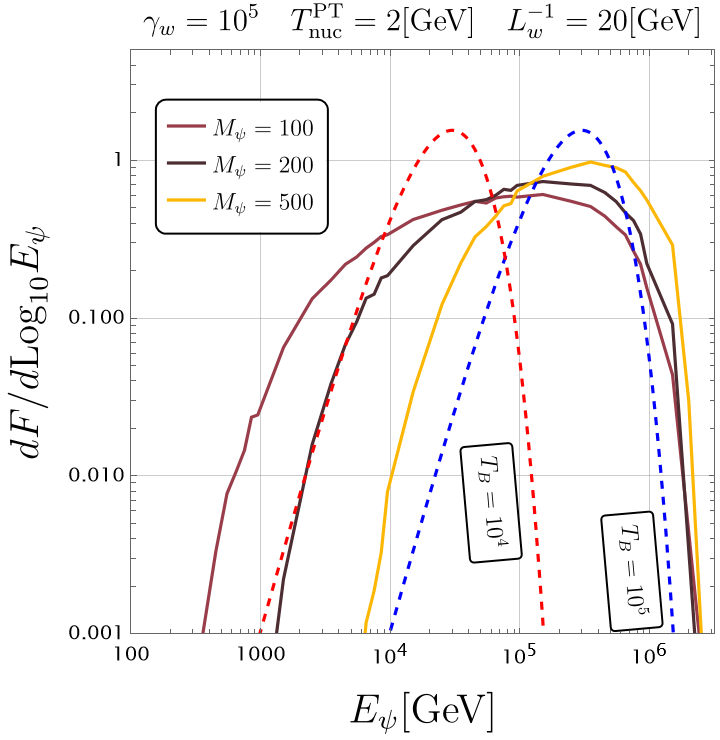}
\includegraphics[scale=0.33]{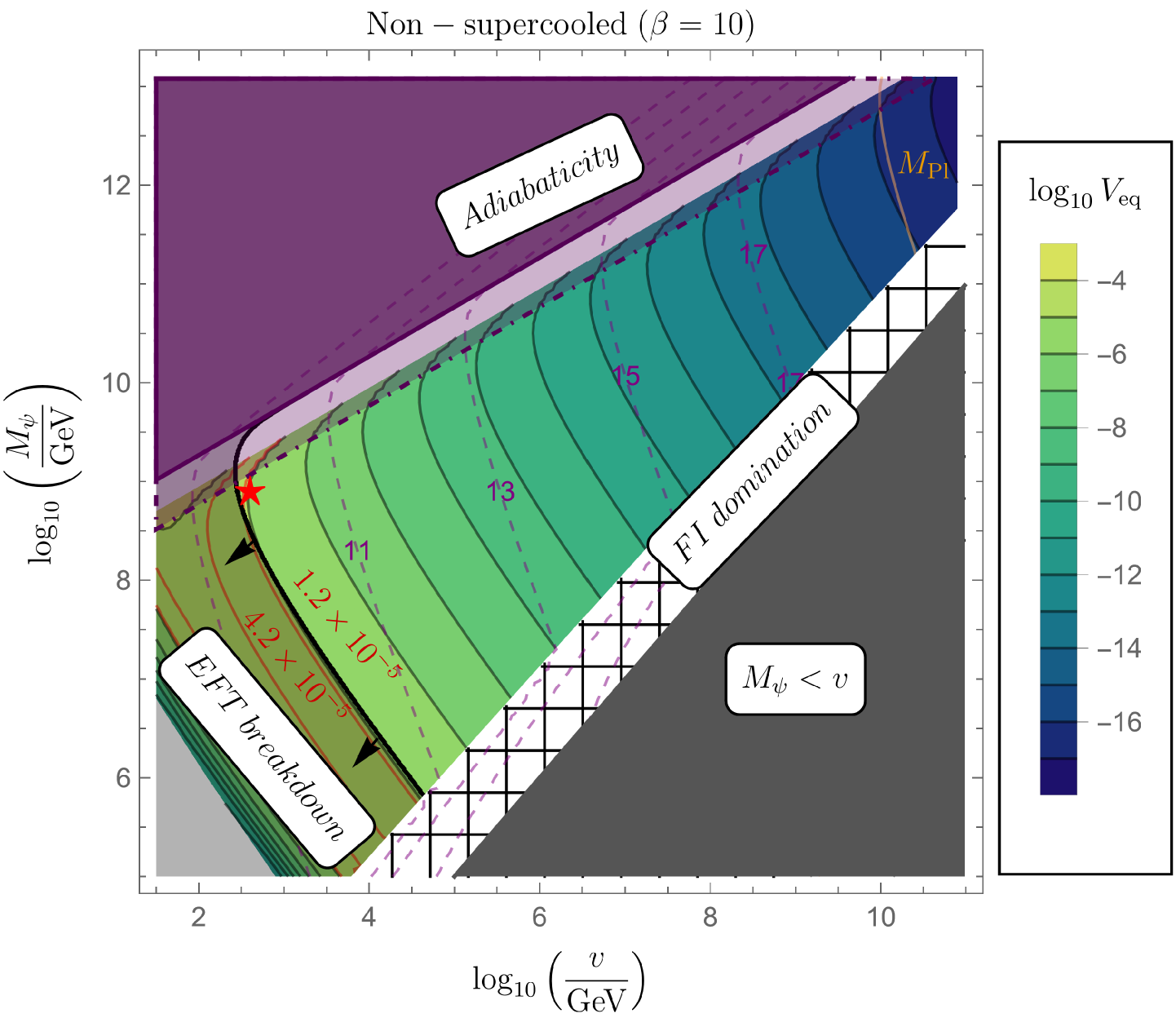}
}
 \caption{Left: Spectrum of the free-streaming fermion after their emission from the bubble wall. Right: Contour plots of $\log_{10}(V_{\rm eq})$ for the operator $\frac{h^2\bar{\psi}\psi }{\Lambda}$ and for $\beta =10$ and $T_{\rm nuc} \sim v$, while $T_{\rm reh}=v$.
Purple dashed lines indicate the isocontours of the UV cutoff, 
$\text{log}
_{10}\left(\frac{\Lambda}{\text{GeV}}\right)$.
The shaded area to the left of the  solid black line excludes the region where the EFT 
analysis breaks down, i.e. $2\gamma v T_{\rm nuc} > \Lambda^2$.
The dark gray region in the lower right of each plot indicates $v>M_\psi$. 
The dark purple area indicates the region where the non-adiabatic condition 
is not satisfied $\gamma_w T v \leq 2 M_{\psi}^2$. The light purple area denotes the region defined by the conditions  $ M_\psi^2\in\gamma_w T_{\rm nuc} v \times [0.05,0.5]$.
In the white hatched regions, FI is the dominant process for DM production. The red lines represent the current and future experimental bounds on the velocity, and the red star marks the specific points we study in the main text. Light grey area in the lower right corner of the plot indicates the DM in the thermal equilibrium.  Adapted from \cite{Azatov:2024crd}.}
 \label{fig:para_space_dim5}
\end{figure}

\begin{figure}[!h]
\hspace{-0.9cm} 
\centering{
\includegraphics[scale=0.48]{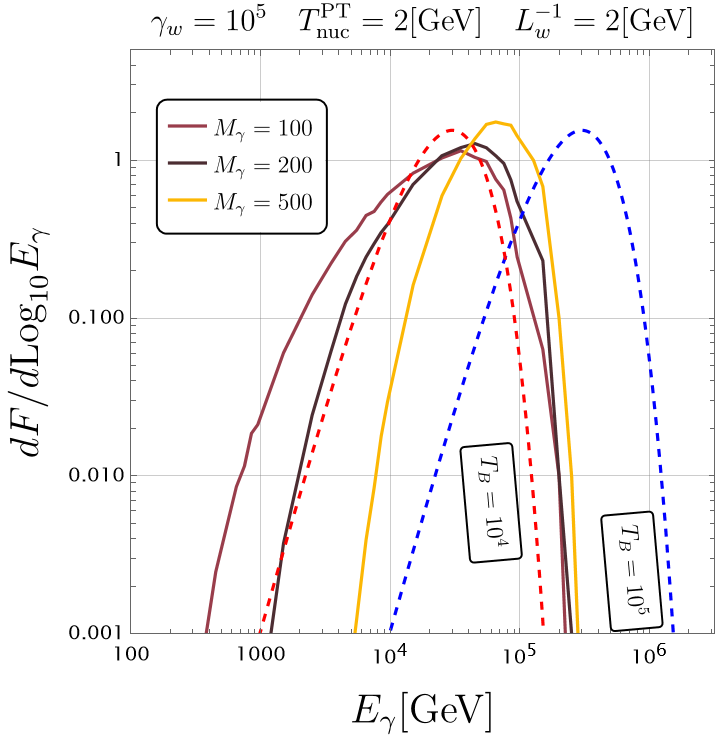}
\includegraphics[scale=0.45]{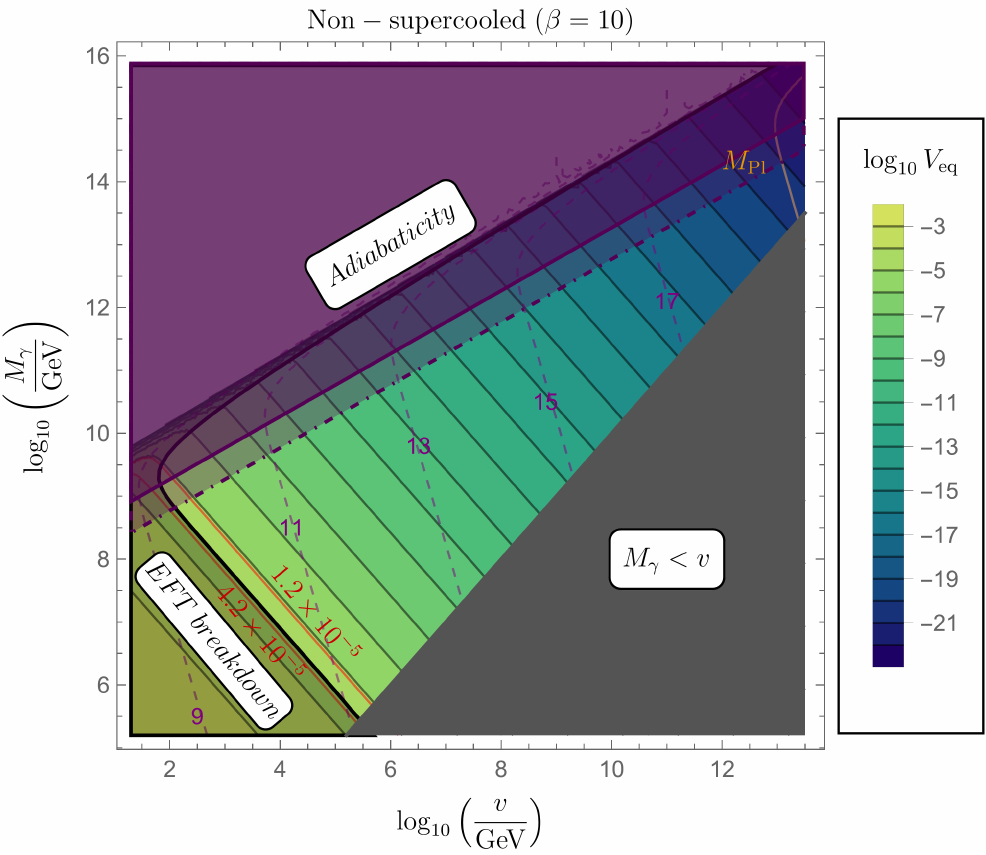}
}
 \caption{Same Figure than \ref{fig:para_space_dim5} but for the case of the dark photon production. Adapted from \cite{Azatov:2021ifm}.}
 \label{fig:para_space_dim6}
\end{figure}

This condition will be shown in the Figures \ref{fig:para_space_dim5}, \ref{fig:para_space_dim6} via a gray region called ``EFT breakdown''.

\paragraph{Results}

We then present the main results of this study. We show the region of the parameter space which can explain the observed DM abundance in Fig.\ref{fig:para_space_dim5} for the fermion emission, Fig.\ref{fig:para_space_dim6} for the dark photon emission and finally for the gluon emission in Fig. \ref{fig:para_spaceGG}, and a Table \ref{tab:Summary_tab}, which contains a summary of all the analytical results of this study. On Fig \ref{fig:para_space_dim5} and \ref{fig:para_space_dim6}, we present on the left panel the spectrum of the DM at emission and observe on the right panel that in most of the acceptable region of the parameter space, the DM will be free streaming, thus keeping the same spectrum than at emission. On the other hand, Glueball DM are always strongly interacting and can never free-stream, we thus do not show their spectrum at emission.  More details about the computations are provided in \cite{Azatov:2024crd}. A benchmark point with very heavy and warm DM, $\beta= 10$ and positive prospects of detection of the gravitational wave signal is 
\begin{align}
    &v= 400 \text{ GeV}, \quad M_\psi = 8\times 10^8 \text{ GeV}\\ 
    \nonumber
    &\Lambda= 6.3 \times 10^9\text{ GeV}, \quad \gamma_w  = 1.7\times 10^{14}, \quad V_{\rm eq} = 9.5\times 10^{-6} \, .
\end{align}

This point is represented on Fig. \ref{fig:para_space_dim5} by a red star and might be soon observable by structure formation probes like Lyman-$\alpha$ and 21 cm\cite{Sitwell:2013fpa,Munoz:2019hjh} or sub-halo count~\cite{LSSTDarkMatterGroup:2019mwo}.

\section{Conclusion}

First order phase transitions (FOPTs), when they are strong and long enough, are powerful emitters of gravitational waves. With the forthcoming GW observers like LISA and ET, this opens the possibility of probing FOPTs in the early universe. However those strong and long phase transitions, possibly supercooled, are naturally expected to have a crucial impact on other processes and properties of the early universe plasma, like DM abundance, baryon and lepton number, magnetic fields. In this talk, we have studied the possibility of producing the observed abundance of DM thanks to a FOPT. In this mechanism, the DM is produced via the interaction between the relativistic bubble wall and the plasma and results in possibly very heavy and boosted DM relics. 

We studied first the emission of scalar dark matter via a renormalisable portal between the dark matter $\phi$ and the transition sector $h^2 \phi^2$, and then moved to study a similar mechanism for a secluded sector $h^2 \psi \bar \psi/\Lambda $, 
$h^2 F_{\mu \nu} F^{\mu \nu}/\Lambda^2$ and $h^2 G_{\mu \nu} G^{\mu \nu}/\Lambda^2$. 
We showed that the DM could be several orders of magnitude more massive than the scale of the FOPT $v$. Since it is produced via the bubble wall, the DM is initially extremely boosted and can remain relativistic for longer time than its thermal counterparts. This opened the possibility of heavy and warm dark matter, while still explaining the observed abundance. As a consequence, this study goes in the direction of relating the production of dark matter with the physics of FOPTs and making the mechanism generating DM testable.

\section*{Acknowledgments}

MV is supported by the ``Excellence of Science - EOS'' - be.h project n.30820817, and by the Strategic Research Program High-Energy Physics of the Vrije Universiteit Brussel.

\bibliographystyle{JHEP}
{\footnotesize
\bibliography{biblio}}

\end{document}